\documentstyle[preprint,aps]{revtex}
\begin{document}
\draft
%\preprint{UNC-CH-CM/4-93}
\title{
Metal-Insulator Transitions in Degenerate Hubbard Models \\
and A$_x$C$_{60}$
}
\author{Jian Ping Lu}
\address{
Department of Physics and Astronomy, \\
University of North Carolina at Chapel Hill,
Chapel Hill, North Carolina 27599
}
\date{\today}
\maketitle

\begin{abstract}
Mott-Hubbard metal-insulator transitions in
$N$-fold degenerate Hubbard models are studied
within the Gutzwiller approximation.
For any rational filling with $x$ (integer) electrons per site
it is found that
metal-insulator transition occurs
at a critical correlation energy
$U_c(N,x)=U_c(N,2N-x)=\gamma(N,x)|\bar{\epsilon}(N,x)|$, where
$\bar{\epsilon}$ is the band
energy per particle for the uncorrelated
Fermi-liquid state and $\gamma(N,x)$ is a geometric
factor which increases linearly with $x$.
We propose that the alkali metal doped
fullerides $A_xC_{60}$ can be described
by a 3-fold degenerate Hubbard model. Using
the current estimate of band width and correlation
energy this implies that most of ${\rm A_xC_{60}}$, at integer $x$,
are Mott-Hubbard insulators and ${\rm A_3C_{60}}$ is a strongly
correlated metal.
\end{abstract}

\pacs{PACS numbers: 71.10.+x,71.30.+h,74.70.W}

%\twocolumn

The discovery of superconductivity in ${\rm A_3C_{60}}$\cite{a3c60}
has spurred great interest
in alkali metal doped fullerides\cite{review}.
Beside ${\rm A_3C_{60}}$, stable phases such as
$Rb_1{\rm C_{60}}$, $Na_2{\rm C_{60}}$, $K_4{\rm C_{60}}$ were
synthesized \cite{review,murphy}.
One unusual property is that except ${\rm A_3C_{60}}$ all integer $x$ phases
are found to behave like insulators\cite{weaver}.
This contradicts the band structure calculations
which imply that all of them are metals due to the 3-fold degeneracy
of the $t_{1u}$ molecular orbitals which forming the conduction
bands\cite{band}.
In this letter we show that
the strong (compared with the band width)
intramolecular electron-electron
correlation is responsible for this unusual property.
The results we have obtained also shed light on the instability of
the non-integer $x$ phases\cite{a1c60}.

The existence of strong correlation in pure ${\rm C_{60}}$
is supported by spectroscopy experiments.
Photoemission shows an insulating gap of 2.6eV, while
the photo-conductivity and absorption indicate
excitation at 1.6eV. This discrepancy is interpreted as due
to strong correlation which results in a
large excitonic binding energy.
The estimated correlation energy $U \sim 1eV$\cite{weaver,lof,uvalue}
is much larger than the conduction band width
$W \sim 0.2 - 0.4eV$\cite{band,gelfand}.
Thus, it has been suggested that
that ${\rm A_3C_{60}}$ is a Mott-Hubbard insulator
and the superconducting phase is non stoichiometric\cite{lof}.
However, structural, transport and spectroscopic measurements
show that the superconducting phase is stoichiometric
and there is no evidence of insulating behavior in
${\rm A_3C_{60}}$. Even more interesting is that for $x\ne3$
integer stoichiometric phases
no metallic behavior have been observed so far.
Therefore neither a simple Hubbard model which prefers
insulating at half filling ($x=3$),
nor the simple band filling model which predict metallic
behavior for all phases, can explain the
unusual metal-insulator transitions observed.

Clearly the 3-fold degeneracy of the conduction band
can not be neglected. This leads us
to study the general $N$-fold degenerate Hubbard model
at rational fillings. We find
that the unusual metal-insulator
transitions observed can be understood in term of Mott-Hubbard transition
in the degenerate Hubbard model.
It is found that for a general $N$-fold degenerate Hubbard model
at rational filling $x/2N$, where the average number of electrons
per site ($x$) is an integer, the metal-insulator transition occurs
at a critical $U_c$ which increases with both $x$ and $N$.
$U_c$ is found to be the largest at half filling $x=N$ for a given
degeneracy $N$ (except $N=2$). Therefore it is possible that the system
is a metal at half filling while insulating away from it.
Our results lay a solid theoretical foundation for
the interpretation that ${\rm A_xC_{60}}$ are either Mott-Hubbard insulators
or strongly correlated metals and provide a rationale to
understand the unusual metal-insulator transitions in this
family of materials and molecular metals in general.

Consider the general $N$-fold degenerate Hubbard model
with the correlation energy $U$ independent of orbitals
and spins
\begin{equation}
H= \sum_{i,j,\alpha,\beta} t_{i,j}^{\alpha,\beta}
c_{j,\beta}^{+} c_{i,\alpha}
+ \frac{U}{2} \sum_{i,\alpha \ne \beta} n_{i,\alpha}n_{i,\beta} \, ,
\end{equation}
where $\alpha=(r,\sigma)$ include
both spin ($\sigma$) and orbital ($r$) indices;
$n_{i,\alpha}=c_{i,\alpha}^{+}c_{i,\alpha}$ are
number operators. Let $L$ be the size of the lattice
and $M$ be the total number of electrons.
By rational filling we mean that the average
number of electrons per site $x=M/L$ is an integer.
For such a filling there exists a well defined insulating state
where there are exactly $x$ electrons localized at each site.
Obviously for a sufficiently large $U$ hopping is forbidden, and the
the ground state is insulating.
As $U$ decreases a metal-insulator transition, at a certain
critical $U_c$, is expected.
For the case of the non-degenerate Hubbard model the
only rational filling is the half filling ($N=x=1$);
in this case it is well known that the ground state
at large U is an ordered magnetic insulator\cite{hubbard}.
For the degenerate Hubbard model, in general, the insulating
state could also be ordered. However, we will consider the
paramagnetic (or disordered) insulating state only because
our primary interest is $A_xC_{60}$,
where the lattice is non bi-partile and large amount of
intrinsic disorders are known to exist\cite{gelfand,disorder}.

\widetext

\narrowtext

In the insulating state,the kinetic energy is zero and
the total energy per site $E_0$ is given by the correlation
energy $PE=\frac{U}{2}x(x-1)$.
Imagining a situation very close to the metal-insulation transition
such that only one site has $x+1$ electrons,
the correlation costs U while the kinetic
energy gained for the excitation is of the order $W$.
Since there are $x$ electrons per site, there are
$x$ possible ways of making such an excitation. Therefore
one might expect $U_c(x) \sim xW$\cite{note1}.

In the case of the non-degenerate Hubbard
model at half filling
the rigorous result $U_c=8|\bar{\epsilon}|=2W$
was obtained by Brinkman and Rice\cite{br} within the
Gutzwiller approximation\cite{gutz}.
The central point of Gutzwiller approximation is
to associate a projection factor $\eta$
with every doubly occupied site, assuming that
the many-body wavefunction
can be written as a superposition of states with different numbers
of doubly occupied sites $\nu$. The optimal $\nu$
is determined variationally
by calculating the expectation value of the Hamiltonian.
In the thermodynamic limit
the summation over $\nu$ is dominated by the optimal $\nu$ term,
then the kinetic and potential energies can be calculated by counting
the number of configurations which contribute.

We have carried out similar calculations rigorously
for the general $N$-fold degenerate Hubbard model
in a limit close to the metal-insulator
transition. The details of counting are rather
tedious and will be published
elsewhere\cite{jpl}.
Here we will simply state assumptions and results
and discuss their implications for ${\rm A_xC_{60}}$.

Let $x$ be the average
number of electrons per site and $L$ be the
total number of sites.
We assume that there is complete
permutation symmetry between
all orbitals and spins, so the number of
electrons occupying each $\alpha=(r,\sigma)$ state
is $m=\frac{xL}{2N}$.
Near the metal-insulator
transition the probability that a site is occupied
by more than $x+1$ electrons is very small as it costs
$2U$ or more energy. Thus we assume that each site
can only be ``empty'' ($x-1$ electrons),
``singly-occupied'' ($x$ electrons)
or ``doubly-occupied'' ($x+1$ electrons).
Let $2N\nu \ll L$ be the total number of doubly occupied
sites, then by symmetry the number of empty sites is also
$2N\nu$. Every doubly occupied site
costs a correlation energy $U$ with respect to
the insulator state where there are exactly
$x$ electrons on every site.
The Gutzwiller wavefunction is constructed from the
uncorrelated Slater determinant
$|SL>=|\{k_1,\alpha_1, ...., k_{n},\alpha_{n}\}>$ by projecting out
doubly occupied states with a weighting factor $\eta$
\begin{equation}
|\phi> = \prod_{i,\alpha,\beta}
(1-\eta c_{i,\alpha}^{+}c_{i,\beta}^{+}) |SL>  \, .
\end{equation}
This wave function is a linear combination of a large number
of states with different $\nu$. Following the original
calculation of Gutzwiller, the expectation value
of the Hamiltonian is dominated by the state with the optimal
$\nu$ which is to be determined variationally. After
a lengthy derivation the average energy
per particle with respect to the paramagnetic insulating
state is found to be\cite{jpl}
\begin{equation}
E(N,x) = Q(N,x,\nu,m)\bar{\epsilon}(x)+\frac{\nu}{m}U \; ,
\end{equation}
where $\bar{\epsilon}(x)$
is the kinetic (band) energy per particle in the uncorrelated state
with the center of the band chosen to be zero,
$\bar{\epsilon}(x=2N)=0$.
The quotient $Q$, which reflects the reduction of the hopping term
in the correlated state\cite{gutz,jpl}, is given by
\begin{equation}
Q(N,x) = \frac{\alpha_{N,x}\nu(m-2x\nu)}{m^2}
\left(1+\eta\frac{(m-2x\nu)}{\alpha_{N,x}\nu}\right)^2 \, ,
\end{equation}
where
\begin{equation}
\alpha_{N,x} = \left\{ \begin{array}{ll}
	\frac{2Nx}{2N-1} & \mbox{\hspace{0.3in} if $x=1$ or $2N-1$} \\
	x		& \mbox{\hspace{0.3in} if $2\leq x\leq 2N-2$} \; ,
	\end{array}
	\right.
\end{equation}
and the projection parameter $\eta$ is
\begin{equation}
\eta = \left\{ \begin{array}{ll}
	\frac{2x\nu}{m-2x\nu} \sqrt{\frac{N}{2N-1}} & \mbox{\hspace{0.3in} if $x=1$ or
$2N-1$} \\
	\frac{x\nu}{m-2x\nu} & \mbox{\hspace{0.3in} if $2\leq x\leq 2N-2$} \; .
	\end{array}
	\right.
\end{equation}
Substituting eqs.4-6 into Eq. (3) and
minimizing the energy with respect to $\bar{\nu}=\nu/m$ leads
to
\begin{equation}
\bar{\nu} = \frac{1}{4x}(1-\frac{U}{U_c})
\end{equation}
and the energy per particle
\begin{equation}
E_0 = \frac{\gamma(N,x)\bar{\epsilon}}{8x} (1-\frac{U}{U_c})^2 \; ,
\end{equation}
where $U$ is given by
\widetext
\begin{eqnarray}
U_c(N,x) &=& \gamma(N,x)|\bar{\epsilon}(N,x)|
	= \left\{ \begin{array}{ll}
	\frac{2Nx}{2N-1} {\left( 1+\sqrt{\frac{2N-1}{N}} \right)}^2
		|\bar{\epsilon}(N,x)|
	&  \mbox{\hspace{0.3in} if $x=1$ or $2N-1$} \\
	4x|\bar{\epsilon}(N,x)| & \mbox{\hspace{0.3in} if $2\leq x\leq 2N-2$}\; .
	\end{array}
	\right.
\end{eqnarray}
\narrowtext
As $U$ increases toward $U_c$, the number of doubly occupied
states approach zero.
Thus the Mott-Hubbard transition occurs at $U_c$.

Eq.9 is the main result of this paper.
Several points are worth mentioning. a)
The particle-hole symmetry is preserved if
$|\bar{\epsilon}(N,x)|=\frac{x}{2N-x}|\bar{\epsilon}(N,2N-x)|$;
this is expected because the starting Hamiltonian
Eq. (1) contains particle-hole symmetry.
b) For $N=1$ the only rational filling is half filling x=1,
the Brinkman-Rice result $U_c=8|\bar{\epsilon}|$ is recovered.
c) Regardless of the band structure, $U_c$ generally increases
with $x$ reaching the maximum at half filling $x=N$
($N=2$ is an exception); thus, for a degenerate Hubbard system
it is more difficult for it to become a Mott insulator.
d) The above results apply only to integer $x$. For non-integer
$x$, if $U$ is larger than $U_c$, there will be tendency
for the system to phase separate into integer
phases to lower the total energy.
This may explain
the experimental observation that non-integer $x$ phases1
are not stable
${\rm A_xC_{60}}$ where the alkali ions can
segregate together with electrons which screen
out the long range Coulomb repulsion.
Details of this issue will be explored
elsewhere.

To make a quantitative estimate, one needs to know
the band structure. In the case of a flat band with the bandwidth $W$,
$\bar{\epsilon}(N,x)=\frac{x-2N}{4N}W$.
One obtains the simple result $U_c(N,x)=\frac{x(2N-x)}{N}W$
($x\neq 1,2N-1$).
This agrees qualitatively with the simple argument
discussed earlier. Fig.1 shows plots of $U_c/W$ vs $x/2N$
for several $N$ values.

Now let us turn to the specific application of the above results
to the metal-insulator
transitions in $A_xC_{60}$. It is known that each alkali metal donates
one electron and that the conduction
band is formed by overlap of the 3-fold degenerate
$t_{1u}$ molecular orbitals.
For $x=1,3,4$ structures are known to be rhombohedral,
face-center-cubic
and body-center-tetragonal, respectively\cite{murphy}. For all these structures
LDA {\it ab initio} calculations suggest that all of them are
metals\cite{band}.
Experimentally, except $x=3$, all phases with integer $x$
are found to be insulating.
It has also been shown that the band structure can be accurately
represented by a 3-band tight-binding model, in particular
the density of states was shown to be approximately flat
due to the intrinsic orientational disorder\cite{gelfand}.
The band width $W$ determined from both
experiments\cite{exw} and calculations\cite{band,gelfand}
is very small,
$W \sim 0.2 -0.4 eV$. On the other hand,
spectroscopic studies\cite{weaver,lof}
and theoretical calculations\cite{uvalue}
suggest that the intramolecular electron correlation
energy is around $U \sim 1eV$.
The value of $U$ is expected to remain unchanged with
doping because the screening is provided by the large
number of molecular orbitals above $t_{1u}$, which is
not affected by the doping.
{}From Eq. (9) and assuming a flat band one obtains
$U_c(x,N=3)/W=2.6,2.67,3$ respectively for $x=1,2,3$.
Thus if the parameters are such that
$2.67W<U<3W$, which is where the current
best estimate of $W$ and $U$ fall, then the whole family
of ${\rm A_xC_{60}}$ for $x=1,2,4,5$ are Mott-Hubbard insulators while $x=3$
could be a strongly correlated metal. Of course we caution that the $U_c$
calculated here only represents the lower bound; however,
we expect the qualitative results
$U_c(N,x)\leq U_c(N,N)$, $U_c(N,N) \sim NW$
will hold for the exact $U_c$.

In conclusion we have studied Mott-Hubbard transitions in the
$N$-fold degenerate Hubbard model within the Gutzwiller approximation.
It is shown that for any integer number of electrons per site there
exists a critical correlation energy $U_c$ above
which the system is a Mott-Hubbard insulator.
$U_c$ is found to be sensitive to both the degeneracy and filling.
We propose that the family of materials ${\rm A_xC_{60}}$
can be described by a 3-fold degenerate Hubbard model.
With reasonable estimates for the band width and the intramolecular
correlation energy, we show that it is possible
that for most integer phases ( $x=1,2,4,5$)
the materials are Mott-Hubbard insulators
and ${\rm A_3C_{60}}$ is likely a strongly correlated metal
despite the fact that $U$ is several times of the band width.

\acknowledgments
I am grateful to M.P. Gelfand for many discussions, and collaborations
on related subjects. I thank A. Hebard,
J. Fisher and J. Weaver for providing
experimental results prior to publications. Discussions with
Q. Si are acknowledged.
This work is supported by The University Research Council of
the University of North Carolina at Chapel Hill.

\begin{figure}

\caption{Phase diagrams for Mott-Hubbard metal-insulator transitions
in $N$-fold degenerate Hubbard models. Shown are results
for $N=2,3,4,5$. For $U>U_c(N,x)$ the system
is a Mott-Hubbard insulator.
Note that only the points, corresponding to
integer ($x$) number of electrons per site, are meaningful.}

\end{figure}


\begin{references}
\bibitem
{note1}
Notice that in the Hartree mean field approximation
it is straightforward to show that
the expectation of Hubbard U term is
$x\frac{2N-1}{4N}U$ per particle, while
the average kinetic energy is
$\epsilon(x)=\frac{x-2N}{4N}W$ for a flat band with the
band width $W$.
Thus, the energy gain with respect to the disordered
insulating state is
$\Delta E=\frac{x-2N}{4N}(W-U)$. Thus
$U_c=W$ independent of $N$ and $x$.
\end{references}
\end{document}